\documentclass{elsart}
\usepackage{amsmath,epsfig,amsfonts,euscript,amssymb,latexsym}

\newcommand{\C}{\mathbb{C}}

\newcommand{\sr}{\mathcal{R}}
\newcommand{\real}{\mathrm{Re}\,}
\newcommand{\imag}{\mathrm{Im}\,}
\newcommand{\tr}{\text{\rm tr}\,}

\renewcommand{\labelenumi}{(\arabic{enumi})}

\begin{document}
\hfill {\small\it Accepted for publication in Physica D}
\begin{frontmatter}

\title{
Maximum Performance at Minimum Cost\\
in Network Synchronization
}

\author[SMU]{Takashi Nishikawa\thanksref{corres_auth}} and
\author[NWU]{Adilson E. Motter}

\address[SMU]{
Department of Mathematics, Southern Methodist University, \\
Dallas, TX 75275, USA}

\address[NWU]{
Department of Physics and Astronomy, Northwestern University, \\
Evanston, IL 60208, USA}

\thanks[corres_auth]{Corresponding author: {\tt tnishi@smu.edu}, +1-214-768-2514 (TEL), +1-214-768-2355 (FAX)}
\thanks{NOTICE: this is the author's version of a work that was accepted for publication in Physica D: Nonlinear Phenomena. Changes resulting from
the publishing process, such as peer review, editing, corrections, structural formatting, and other quality control mechanisms may not be reflected
in this document. Changes may have been made to this work since it was submitted for publication.}

\date{Sept. 24, 2006}

\begin{abstract}
We consider two optimization problems on synchronization of oscillator networks:  maximization of synchronizability and minimization of synchronization cost.  
We first develop an extension of the well-known master stability framework to the case of non-diagonalizable Laplacian matrices.
We then show that the solution sets of the two optimization problems coincide and are simultaneously characterized by a simple condition on the Laplacian eigenvalues.
Among the optimal networks, we identify a subclass of {\em hierarchical} networks, characterized by the absence of feedback loops and the normalization of inputs.
We show that most optimal networks are {\em directed} and {\em non-diagonalizable}, necessitating the extension of the framework.
We also show how {\em oriented spanning trees} can be used to explicitly and systematically construct optimal networks under network topological constraints.   
Our results may provide insights into the evolutionary origin of structures in complex networks for which synchronization plays a significant role.
\end{abstract}

\begin{keyword}
  complex networks \sep synchronization \sep optimization
  \PACS 05.45.Xt \sep 89.75.-k \sep 87.18.Sn
\end{keyword}

\end{frontmatter}

\section{Introduction}

Synchronization has recently been attracting many researchers' attention as a simple yet interesting example of collective behavior on complex networks~\cite{Zhou:2006b,Li:2006,Stilwell:2006,Arenas:2006,Wiley:2006,Park2006,Boccaletti2006,Hasegawa2006,Donetti:2005,Morelli2005,Restrepo2005,Oh:2005,Grinstein:2005,Wu:2005,Masoller:2005,Jalan:2005,Lind2004,Atay:2004,Belykh:2004,Timme:2004,Wang,Donetti}. Of particular interest is how the networks' ability to synchronize depends on various structural parameters of the networks, such as average node-to-node distance~\cite{barahona2002,Nishikawa2003}, clustering coefficient~\cite{McGraw2005}, degree distribution~\cite{Nishikawa2003,Kocarev2005}, and weight distribution~\cite{Motter2005b,Motter2005c,Motter2005}.  Revealing the precise mechanism for synchronization in relation to the network structure is an important step toward understanding more complex behavior and unveiling the evolutionary origin of real-world networks.

In some naturally evolved networks, such as neuronal and biochemical
networks, there is evidence that synchronized or more general coordinated
behavior may be playing signicant roles in the system's functionality~\cite{Winfree,Rodriguez,Stopfer}.  It appears natural to expect that the
ability of such networks to synchronize their activity has been optimized
to some extent during their evolution. In reality, however, the problem is
likely to be more complicated, since the fitness of the networks could
depend on multiple factors, such as stability, robustness, and
adaptability. Because of the complexity of the problem, researchers have
instead been focusing on simplified yet tractable optimization problems as a
first step toward solving the real problem.
In particular, there have been some efforts to solve the problem of maximizing the synchronizability of oscillator networks~\cite{Donetti:2005,Wang,Donetti,Motter2005b,Motter2005c,Motter2005,Hwang2005}.  The scope of such studies has so far been mainly limited either to numerical investigations or to
single-parameter families of possible networks.

In this paper, we go beyond these restrictions and present rigorous
solutions to the problem of maximizing the network synchronizability,
measured by the range of coupling parameter for which the system
achieves stable synchronization. 
We also consider a more general concept, the \textit{cost} required for stable synchronization, and treat the problem of minimizing it.
Remarkably, we prove
that the solution sets of the two optimization problems coincide and are
completely characterized by a simple condition on the Laplacian
eigenvalues of the network. This spectral characterization, however, does
not provide much intuition about the structure of the optimal networks. To
gain more intuition, we explicitly construct a large subclass of optimal
networks characterized by a hierarchical
structure, in which information can flow only from top to bottom of the
hierarchy, making the network optimal for synchronization.

The theorems in this paper also provide mathematical foundation for the
solutions of a related problem that we presented in our recent publication~\cite{Nishikawa:2006}. That problem was originally motivated by the discovery that random
scale-free and other degree-heterogenous networks are generally difficult
to synchronize~\cite{Nishikawa2003}.  This discovery led to efforts to enhance the
synchronizability of complex networks by introducing directionality and
weight to each link~\cite{Motter2005b,Motter2005c,Motter2005}. 
Underlying such efforts is a problem of
optimization with topological constraints~\cite{Nishikawa:2006}:  given a fixed topology of
allowed interactions, find assignments of weights and directions that
would maximize the range of the coupling parameter for stable
synchronization. Here we prove that a solution can be systematically and
explicitly constructed using \emph{oriented spanning trees} embedded within any
given connected topology of allowed interactions.  The resulting
networks are guaranteed to be optimal with respect to both
synchronizability and synchronization cost.

The problem of optimization under topological constraints is potentially relevant for many real-world networks.
In metabolic networks, for example, the weights and directions of feasible links (metabolic fluxes) are adjusted to optimize fitness, which is likely to account for robustness of synchronized behavior against environmental changes~\cite{fishcer:2005}.  Similar adjustment of weights and directions may enhance neuronal synchronization within a given topology of synaptic connections in the brain.  In designing the interaction scheme for a computer networks, choosing proper weights and directions may optimize the performance of computational tasks based on synchronization of processes~\cite{Korniss:2003}.  The adjustment of flows in power grids and communication patterns
in social organizations are additional examples where directional and weighted patterns may be favored because they can facilitate the synchronized or coordinated behavior on which the functioning of these networks is based.

Our results are based on an extension~\cite{Nishikawa:2006} of a well-known framework for
studying network synchronization~\cite{pecora1998}.
The power of this framework is that it can separate the effect of
the network structure from that of the dynamics of individual nodes.
However, it implicitly assumes that the Laplacian matrix of the network is
diagonalizable, i.e., the dynamics of the network must be decomposable
into independent eigenmodes. In most of the previous works, this
assumption was automatically satisfied, since the main focus was on
symmetrically coupled networks, which are guaranteed to be diagonalizable.
However, the same does not hold true in general when the network is
directed. In fact, we prove the interesting result that most
optimal networks are non-diagonalizable, and thus violate an assumption
of the original framework. 
We show that the stability condition for synchronization is formally the
same for all networks, but the speed at which the system converges toward the synchronized state can be significantly slower when the network is non-diagonalizable.

The technique underlying our extended framework can be regarded as an
example of a methodology for studying complex systems that relies neither
on the eigenmode decomposition nor on any kind of superposition principle,
and is expected to meet applications in various other network phenomena.

The paper is organized as follows.  
In the next section, we present the extended master stability framework.  We state the optimization problems in Section 3 and characterize their solutions in Section 4, 5, and 6.  
In Section 7, we consider the optimization problems with topological constraints. 
Finally, in Section 8, we make concluding remarks on implications of our results and on future directions.

\section{Extension of Master Stability Analysis\label{sec:extension}}

Consider $n$ identical oscillators whose individual dynamics without coupling is governed by $\dot{\mathbf{x}} = \mathbf{F}(\mathbf{x})$, $\mathbf{x}\in I\!\! R^m$. 
Now consider the network of these oscillators interacting through a diffusive-type coupling, {\it i.e.}, oscillator $i$ receives input from oscillator $j$ that is proportional to  $A_{ij}[\mathbf{H}(\mathbf{x}_j) - \mathbf{H}(\mathbf{x}_i)]$, where $A_{ij}$ is a nonnegative constant representing the relative strength of the coupling, and $\mathbf{H}: I\!\! R^m \to I\!\! R^m$ is a general output function.  The interaction is indeed the usual diffusive coupling when $\mathbf{H}(\mathbf{x}) = \mathbf{x}$.
The set of equations governing the dynamics of the system is then
\begin{equation}\label{eqn:orig}
\dot{\mathbf{x}}_i = \mathbf{F}(\mathbf{x}_i) + \sigma \sum_{j=1}^n A_{ij}[\mathbf{H}(\mathbf{x}_j) - \mathbf{H}(\mathbf{x}_i)], \quad i=1,\ldots,n
\end{equation}
where $\sigma$ is the parameter controlling the overall coupling strength.  Note that the system can be regarded as a sum of two distinct components:  the {\em network structure} represented by the adjacency matrix $A=(A_{ij})$ and the {\em dynamical component} represented by the functions $\mathbf{F}$ and $\mathbf{H}$.  
The general method of analysis introduced by Pecora and Carroll~\cite{pecora1998} to study the stability of complete synchronization in Eq.~\eqref{eqn:orig} is based on the diagonalization of its variational equation.  In the following, we extend their analysis to include cases where the diagonalization is not necessarily possible.

We first note that having  diffusive-type coupling guarantees the existence of a completely synchronous state, though it may be unstable.  In fact, given any solution $\mathbf{x}=\mathbf{s}(t)$ of the individual dynamics
$\dot{\mathbf{x}} = \mathbf{F}(\mathbf{x})$, the completely synchronized
solution, defined by $\mathbf{x}_i = \mathbf{s}(t)$, $i=1, \ldots ,n$, is automatically
a solution of the entire system~\eqref{eqn:orig}. 
For notational convenience, we rewrite Eq.~\eqref{eqn:orig} as
\begin{equation}\label{eqn:main}
\dot{\mathbf{x}}_i = \mathbf{F}(\mathbf{x}_i) - \sigma \sum_{j=1}^n L_{ij}\mathbf{H}(\mathbf{x}_j),\end{equation}
where $L = (L_{ij})$ is called the Laplacian matrix of the directed weighted network, defined
by 
\begin{equation}
\begin{split}
L_{ij} &= -A_{ij} \quad \text{if $i \neq j$},\\
\displaystyle L_{ii} &= -\sum_{j\neq i} L_{ij}.
\end{split} 
\end{equation}
Like the adjacency matrix $A$, the Laplacian matrix also contains all information about the network structure and can be regarded as a network analog of the Laplacian operator for diffusive processes on continuous space.
Note that $L$ is not necessarily symmetric because in our general formulation the network is not constrained to
be undirected.

The stability condition can be studied by considering the variational
equation for the synchronous solution $\mathbf{x}_i=\mathbf{s}(t)$ of Eq.~\eqref{eqn:main}:
\begin{equation}
\dot{\mathbf{\xi}}_i = D\mathbf{F}({\bf s})\mathbf{\xi}_i - \sigma \sum_{j=1}^n L_{ij}D\mathbf{H}(\mathbf{s})\mathbf{\xi}_j,
\end{equation}
which can also be written in the matrix form as
\begin{equation}\label{eqn:variational}
\dot{\xi} = D\mathbf{F}(\mathbf{s})\xi - \sigma D\mathbf{H}(\mathbf{s})\xi L^T,
\end{equation}
where $\xi = (\boldsymbol{\xi}_1, \ldots ,\boldsymbol{\xi}_n)$
is the $m \times n$ perturbation matrix, $\boldsymbol{\xi}_i$ is the vector of perturbations to the $i$th oscillator, and $L^T$ denotes the transpose of $L$.  
In the Pecora-Carroll analysis, the assumption that the Laplacian matrix $L$ is diagonalizable was implicitly used to diagonalize the variational equation~\eqref{eqn:variational}.  Here we do not assume the diagonalizability of $L$.  Instead, we utilize the Jordan canonical transformation of $L$.  For any $n \times n$
matrix $L$, there exists an invertible matrix $P$ of generalized eigenvectors of $L$,
which transforms $L$ into Jordan canonical form as
\begin{equation}
\label{eqn:jordan}
P^{-1}LP = J = \begin{pmatrix}
\ 0\  & & &\\
& \ B_1\  & & \\
& &\ \ddots\ & \\
& & &\ B_l\ 
\end{pmatrix},
\end{equation}
where $B_i$'s are blocks of the form
\begin{equation}
B_i = \begin{pmatrix}
\ \lambda\ & & & \\
1 &\ \lambda\ & & \\
& \ddots &\ \ddots\ & \\
& & 1 &\ \lambda\ 
\end{pmatrix}
\end{equation}
and $\lambda$ is one of the (possibly complex) eigenvalues of $L$. 
We note that the Jordan canonical transformation has been used to study the stability of synchronization in specific classes of networked systems~\cite{Lu2006,Lu2004a,Lu2004b}.
By applying the change of variable
$\eta = \xi (P^{-1})^T$ in Eq.~\eqref{eqn:variational}, we get~\cite{Nishikawa:2006}
\begin{equation}
\label{eqn:variational2}
\dot{{\eta}} = D\mathbf{F}(\mathbf{s}){\eta} - \sigma D\mathbf{H}(\mathbf{s}){\eta} J^T.
\end{equation}
Each column of $\eta$, being a linear combination of all $\boldsymbol{\xi}_i$'s, represents in general a mode of perturbation to the entire oscillator network, and not to any particular oscillator.  Thus, the synchronous solution is stable if and only if all of these columns converge to zero under Eq.~\eqref{eqn:variational2}.

\begin{figure}[t]
\begin{center}
\epsfig{figure=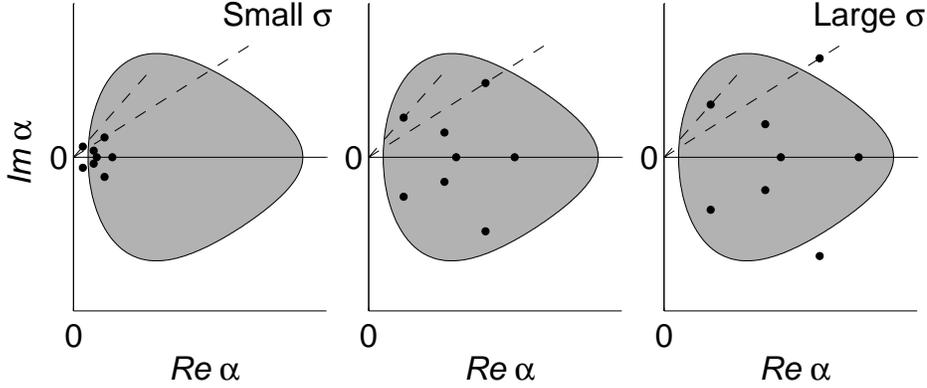,width=0.9\textwidth}
\caption{Schematic illustration of the stability condition~\eqref{eqn:stability} for synchronized state.  The shaded area is the region in the complex plane in which the master stability function $\Lambda(\alpha)$ is negative, and the dots represent $\sigma\lambda_i$ for $i=2,\ldots,n$.  The condition~\eqref{eqn:stability} corresponds to having all the dots in the shaded region.
}
\label{fig:sr}
\end{center}
\end{figure}

Before getting into the general treatment, let us first consider the case where $L$ is diagonalizable.  In this case, the matrix $J$ is a diagonal matrix having the eigenvalues $\lambda_1,\ldots,\lambda_n$ of $L$ along the diagonal.  Thus, the equation for each column of $\eta$ becomes independent of the others and takes the form
\begin{equation}\label{eqn:ms}
\dot{\mathbf{y}} = [D\mathbf{F}(\mathbf{s}) - \alpha D\mathbf{H}(\mathbf{s})] \mathbf{y},
\end{equation}
where $\alpha=\sigma\lambda_i$ when $\mathbf{y}$ represents the $i$th column of $\eta$.  Regarding $\alpha$ as a tunable complex parameter, Eq.~\eqref{eqn:ms} is called a master stability equation and its stability profile as a function of $\alpha$ determines the linear stability of the synchronous solution in systems with various network structures represented by the Laplacian eigenvalues. 
The largest Lyapunov exponent $\Lambda(\alpha)$ for the solution $\mathbf{y}=\mathbf{0}$ is usually used to test the stability and is called a master stability function~\cite{pecora1998}.
The eigenvalues $\lambda_1,\ldots,\lambda_n$ of $L$ can always be arranged so that $0=\lambda_1 \le \real\lambda_2 \le \ldots \le \real\lambda_n$, since $\sum_j L_{ij} = 0$ implies that we always have eigenvalue 0 corresponding to the eigenvector $(1,\ldots,1)^T$, and all eigenvalues are guaranteed to have nonnegative real parts by the Gerschgorin Circle Theorem.  Thus, the condition for the synchronized solution to be linearly stable is
\begin{equation}\label{eqn:stability}
\Lambda(\sigma\lambda_i) < 0,\quad i=2,3,\ldots,n.
\end{equation}
See Fig.~\ref{fig:sr} for a visual demonstration of this stability condition.  Note that $\lambda_1=0$ is excluded from the condition, because $\Lambda(\sigma\lambda_1) = \Lambda(0)$ actually determines the linear stability of the individual solution $\mathbf{s}(t)$ against perturbation [try setting $\alpha=0$ in Eq.~\eqref{eqn:ms}].  It would be positive if $\mathbf{s}(t)$ is chaotic, but it does not affect the stability of synchronization.  In other words, $\Lambda(0)$ corresponds to the stability in the direction parallel to the synchronization manifold defined by $\{ \mathbf{x}_1 = \cdots = \mathbf{x}_n \}$, while $\Lambda(\sigma\lambda_2), \ldots, \Lambda(\sigma\lambda_n)$ correspond to the stability in the directions transversal to the manifold.

Let us now consider the more general case where $L$ is not necessarily diagonalizable.  Each block of the Jordan canonical form corresponds to a subset of these columns in $\eta$, which obeys a subset of equations in
\eqref{eqn:variational2}. For example, if block $B_i$ is $k \times k$, and if the corresponding columns of $\eta$ are denoted by $\boldsymbol{\eta}_{1},\boldsymbol{\eta}_{2},\ldots,\boldsymbol{\eta}_{k}$, then the equations take the form
\begin{align}
\dot{\boldsymbol{\eta}}_{1} &= [D\mathbf{F}(\mathbf{s})
- \alpha D\mathbf{H}(\mathbf{s})]\boldsymbol{\eta}_{1} \label{eqn:eta1}\\
\dot{\boldsymbol{\eta}}_{2} &= [D\mathbf{F}(\mathbf{s}) 
- \alpha D\mathbf{H}(\mathbf{s})]\boldsymbol{\eta}_{2} - \sigma D\mathbf{H}(\mathbf{s})\boldsymbol{\eta}_{1} \label{eqn:eta2}\\
&\cdots\nonumber\\
\dot{\boldsymbol{\eta}}_{k} &= [D\mathbf{F}(\mathbf{s}) 
- \alpha D\mathbf{H}(\mathbf{s})]\boldsymbol{\eta}_{k} - \sigma D\mathbf{H}(\mathbf{s})\boldsymbol{\eta}_{k-1}, \label{eqn:eta3}
\end{align}
where $\alpha=\sigma\lambda$.  Here $\boldsymbol{\eta}_{1},\boldsymbol{\eta}_{2},\ldots,\boldsymbol{\eta}_{k}$ represent the modes of perturbation in the generalized eigenspace associated with eigenvalue $\lambda$.  
Equation~\eqref{eqn:eta1} has exactly the same form as the master stability equation~\eqref{eqn:ms}, so
$\eta_1$ converges exponentially to zero as $t \to \infty$ if and only if $\Lambda(\sigma\lambda) < 0$. 
The condition for Eq.~\eqref{eqn:eta2} to be stable is apparently more involved but
can be formulated
as follows.  Assuming that $\Lambda(\sigma\lambda) < 0$ and that the norm of $D\mathbf{H}(\mathbf{s})$ is bounded, we have that the second term in Eq.~\eqref{eqn:eta2} is exponentially small as well.
Then, the same condition $\Lambda(\sigma\lambda) < 0$ 
guarantees the stabilizing effect of 
both the first and second terms, 
resulting in exponential convergence of $\boldsymbol{\eta}_2$ to zero as $t \to \infty$.
The same argument applied repeatedly shows that $\boldsymbol{\eta}_3,\ldots,\boldsymbol{\eta}_k$ must also converge to zero if
$\Lambda(\sigma\lambda) < 0$.  This shows
that $\Lambda(\sigma\lambda) < 0$ is the condition for the linear
stability of the equations corresponding to each full block $B_i$.  Thus, when all the Jordan blocks are taken into account, we see that the stability condition for the synchronous solution in the general non-diagonalizable case is also given by \eqref{eqn:stability}.

Although the stability condition is the same for both the diagonalizable and
non-diagonalizable cases, it is worthwhile noting that there is a crucial difference.  If $L$ is diagonalizable, then each mode of perturbation is decoupled from others, so the exponential convergence of each column of $\eta$ occurs simultaneously and independently of other columns.  
On the other hand, if $L$ is not diagonalizable, some modes
of perturbation may suffer from a long transient because they may be coupled to other modes associated with the same eigenvalue.  
In Eqs.~\eqref{eqn:eta1}--\eqref{eqn:eta3}, $\boldsymbol{\eta}_2$
may have to wait for $\boldsymbol{\eta}_1$ to get small enough before it can start converging, $\boldsymbol{\eta}_3$ may have to wait for $\boldsymbol{\eta}_2$, and so on, so $\boldsymbol{\eta}_k$
may have to wait for a long time before it starts to converge.  
The larger the size $k$ of the Jordan block, the longer we expect the transient to be.
As a simple example to illustrate this effect, consider a network of linearly coupled phase oscillators described by
\begin{align}
\dot\theta_i &= \omega + \sigma\sum_{j=1}^n A_{ij}(\theta_j - \theta_i), \quad \theta_i \in \mathbb{S}^1, \label{eqn:phase}\\
&= \omega - \sigma\sum_{j=1}^n L_{ij} \theta_j,
\end{align}
where $\mathbb{S}^1$ denotes the unit circle.
In this case, the corresponding variational equation around the synchronized solution $\theta_i = \omega t$, $i=1, \ldots ,n$, is a simple linear system, and so are the corresponding Eqs.~\eqref{eqn:eta1}--\eqref{eqn:eta3}.  Thus, they can be explicitly solved to give
\begin{align}
\eta_1 &=  c_1 e^{-\alpha t}\\
\eta_2 &= (- c_1 \sigma t +c_2) e^{-\alpha t} \\
\eta_3 &= \left[\frac{ c_1 \sigma^2 t^2 }{2} - c_2 \sigma t + c_3 \right]e^{-\alpha t} \\
&\cdots\\
\eta_k &= \left[ \frac{(-1)^{k-1}  c_1 \sigma^{k-1}t^{k-1}}{(k-1)!}
+ \cdots -  c_{k-1} \sigma t + c_k \right]e^{-\alpha t}
\end{align}
where $(c_1,\ldots,c_k)$ is the initial value of the perturbation vector $(\eta_1,\ldots,\eta_k)$.  We can indeed see that the polynomial factors lead to slower convergence for larger $k$ in this example.

In fact, even when $L$ is diagonalizable, we expect to see longer transient as it becomes closer to being non-diagonalizable, {\it i.e.}, some of its eigenvectors become closer to being parallel.  To see this, imagine a small sphere centered at the origin of the space of perturbation matrices $\xi$.  If a pair of eigenvectors are almost parallel, then the matrix $P$ of eigenvectors is close to being singular.  Hence, the sphere is stretched quite a bit along some direction under the transformation $\eta = \xi (P^{-1})^T$.  This implies that small perturbations to the synchronized state in the original coordinates can lead to large perturbations in the eigenvector coordinates.  This in turn leads to relatively long transient, even though the type of convergence remains purely exponential.  This mechanism is at work to some extent for any network with non-orthogonal eigenvectors, but the effect is more prominent if the eigenvectors are closer to being parallel.  In the limit of parallel eigenvectors, $L$ becomes non-diagonalizable, and the convergence becomes qualitatively different, as we saw in the phase oscillator example above.  Thus, we expect to observe relatively long transient not only in a few special networks, but also in many others close to them.

The fact that the stability condition is the same regardless of the diagonalizability of the Laplacian matrix is analogous to the fact that the linear stability condition of a fixed point in a dynamical system is the same regardless of the diagonalizability of the Jacobian.  In both situations, lack of diagonalizability leads to long transient.

\section{Optimization Problems\label{sec:def}}

Here we are interested in two different optimization problems:
\begin{itemize}
\item Which network structure maximizes the synchronizability of the system?
\item Which network structure allows the system to synchronize stably with the minimum possible cost? 
\end{itemize}
To address these optimization problems, we need to precisely define quantities to optimize:  the synchronizability and the synchronization cost of an oscillator network.  
To set the stage for doing this, we first let $\sr$ denote the {\em stability region}, defined as the subset of the complex plane in which the master stability function $\Lambda(\alpha)$ is negative, {\it i.e.}, $\sr = \{ \alpha \in \C \,|\, \Lambda(\alpha) < 0 \}$. Using this notation, the stability condition~\eqref{eqn:stability} can now be written as $\sigma\lambda_2,\ldots,\sigma\lambda_n \in \sr$.  
This shows clearly that there are only two distinct factors that determine the stability of the synchronized solution:  
\begin{enumerate}
\item {\em Network structure}, encoded in the adjacency matrix $A$ and affecting the stability only through the Laplacian eigenvalues $\lambda_2,\ldots,\lambda_n$;
\item {\em Dynamical component}, consisting of the individual dynamics given by $\mathbf{F}$ and $\mathbf{s}$, together with the output signal function $\mathbf{H}$, and affecting the stability only through the properties of the stability region $\sr$.  We denote this component by $(\mathbf{F},\mathbf{H},\mathbf{s})$.
\end{enumerate}
By considering the complex conjugate of Eq.~\eqref{eqn:ms}, we can see that $\sr$ is always symmetric about the real axis.  
In most of the previously studied cases, it has been found~\cite{pecora1998,fink2000} that
\begin{description}
\item[(A-I)] $\sr$ is a convex subset of $\C$.
\end{description}
In addition, for a large class of systems in which the dynamics of each oscillator is chaotic, it has also been found~\cite{Atay:2004,Motter2005c,pecora1998,fink2000} that
\begin{description}
\item[(A-II)] $0 < \alpha_1 < \alpha_2 < \infty$,
\end{description}
where we define
\begin{align}
\alpha_1 = \inf \{\real\alpha \,|\, \alpha \in \sr \},\quad
\alpha_2 = \sup \{\real\alpha \,|\, \alpha \in \sr \}.
\end{align}

The assumptions (A-I) and (A-II) on the stability region imply 
that  the set of overall coupling strength $\sigma$ for which complete synchronization is stable,
\begin{equation}
\begin{split}
I_{\rm sync} &= \{ \sigma \,|\, \Lambda(\sigma\lambda_i) < 0,\; i=2,\ldots,n \}\\
&= \{ \sigma \,|\, \sigma\lambda_2,\ldots,\sigma\lambda_n \in \sr \},
\end{split}
\end{equation}
is either an empty set or a finite interval with endpoints at $\sigma_{\min}$ and $\sigma_{\max}$, where $0 < \sigma_{\min} \le \sigma_{\max} < \infty$.  In the case of a finite interval, this can be physically interpreted as follows.   The completely synchronized state of the network is unstable for small enough values of $\sigma$, but as $\sigma$ is increased, it becomes stable at a lower threshold $\sigma_{\min}$ and then becomes unstable again above an upper threshold $\sigma_{\max}$.   
Thus, the relative width of this interval defined by 
\begin{equation}
S = \frac{\sigma_{\max}}{\sigma_{\min}}
\end{equation}
provides a natural and convenient measure of how easy it is for the network to synchronize, {\it i.e.}, the {\em synchronizability} of a network of coupled chaotic oscillators:  larger values of $S$ correspond to more synchronizable networks.  
In this paper, we consider only those systems with $\sr$ satisfying properties (A-I) and (A-II) to ensure that $S$ is well-defined.

\begin{figure}[t]
\begin{center}
\epsfig{file=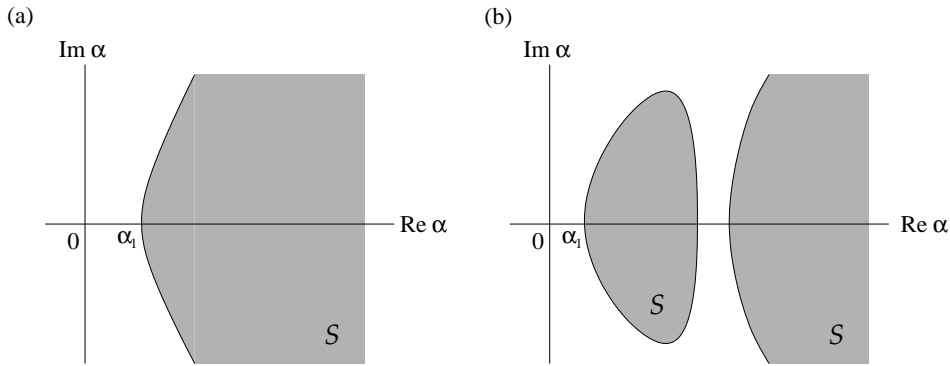,width=0.9\textwidth}
\caption{Examples of stability region $\sr$ for systems with (a) no short-wavelength bifurcation or with (b) intermediate-wavelength bifurcation.}
\label{fig:sr2}
\end{center}
\end{figure}

The {\em synchronization cost} $C$ of a network is defined~\cite{Motter2005b,Motter2005c} as the sum of the total input strength of all nodes at the lower synchronization threshold $\sigma_{\min}$:
\begin{equation}
C = \sigma_{\min}\sum_{i,j=1}^n A_{ij}.
\end{equation}
We define $S=0$ and $C=\infty$ when $I_{\rm sync}$ is empty because this simply means that stable synchronization is impossible.  
Note that while $S$ is guaranteed to give meaningful values only when (A-I) and (A-II) are satisfied, $C$ is meaningful without any assumptions on the stability region.
This means that the notion of synchronization cost applies to a wider range of systems, including those with no short-wavelength bifurcation~\cite{Belykh:2004,Heagy} or with intermediate-wavelength bifurcation.
See Fig.~\ref{fig:sr2} for examples of the stability region for such systems.
The results presented below for the synchronization cost would hold even when these assumptions are weakened, but here we assume (A-I) and (A-II) for simplicity.

With $S$ and $C$ defined, the two optimization problems under consideration can now be formulated precisely:  
\begin{itemize}
\item Which network structure $A$ has the property that it maximizes the synchronizability $S$ for \textit{any} given dynamical component $(\mathbf{F},\mathbf{H},\mathbf{s})$ with $\sr$ satisfying (A-I) and (A-II)?
\item Which network structure $A$ has the property that it minimizes the synchronization cost $C$ for \textit{any} given dynamical component $(\mathbf{F},\mathbf{H},\mathbf{s})$ with $\sr$ satisfying (A-I) and (A-II)?
\end{itemize}
By requiring the optimality for the entire class of dynamical components, we are defining the optimality of a network structure, independently of the dynamical component of any specific system.
Note that both $S$ and $C$ are invariant under re-scaling of $A$, and thus only the relative distribution of the individual coupling strength $A_{ij}$ is important for the optimization problems.

\section{Optimal Network Structures\label{sec:optimal}}

In this section, we give a complete characterization of the class of solutions to the two optimization problems introduced in the previous section.  We start by studying the properties of the quantities to be optimized.  Even though the synchronizability $S$ and the synchronization cost $C$ generally depend on both the dynamical component encoded in $\sr$ and the network structure encoded in $\lambda_2, \ldots ,\lambda_n$, it turns out that they are both bounded by a quantity that depends only on $\sr$ (and also on $n$ in the case of $C$). 
\begin{thm}\label{thm:bounds}
For any stability region $\sr$ satisfying {\rm (A-II)}, we have
\begin{equation}
S \le \frac{\alpha_2}{\alpha_1},
\quad C \ge \alpha_1 (n-1).
\end{equation}
\end{thm}
\begin{pf}
Let the Laplacian eigenvalues be ordered so that  $0=\lambda_1 \le \real\lambda_2 \le \dotsb \le \real\lambda_n$.  If $\real\lambda_2 = 0$, then $I_{\rm sync}$ is empty, and the inequalities are clearly satisfied since $S=0$ and $C=\infty$.  Thus, we are left with proving the inequalities in the case $0 < \real\lambda_2 \le \dotsb \le \real\lambda_n$.

Suppose $\sigma \in I_{\rm sync}$.  Then, since $\sigma\lambda_i \in \sr$ for $i=2,\ldots, n$, we have
\begin{equation}
\alpha_1 \le\real(\sigma\lambda_i) \le \alpha_2,
\end{equation}
and hence
\begin{equation}
\frac{\alpha_1}{\real\lambda_i}  \le \sigma \le \frac{\alpha_2}{\real\lambda_i}
\end{equation}
for $i=2,\ldots, n$.  In particular, this implies that
\begin{equation}
\frac{\alpha_1}{\real\lambda_2}  \le \sigma \le \frac{\alpha_2}{\real\lambda_n}.
\end{equation}
This holds for any $\sigma \in I_{\rm sync}$, so by the definition of $\sigma_{\min}$ and $\sigma_{\max}$, we have
\begin{align}
\frac{\alpha_1}{\real\lambda_2} \le \sigma_{\min} \text{ and }
\sigma_{\max} \le  \frac{\alpha_2}{\real\lambda_n}.
\end{align}
Therefore,
\begin{equation}\label{eqn:bound}
S = \frac{\sigma_{\max}}{\sigma_{\min}} \le 
\frac{\alpha_2}{\alpha_1} \cdot \frac{\real\lambda_2}{\real\lambda_n}
\le \frac{\alpha_2}{\alpha_1},
\end{equation}
and
\begin{align}
C &= \sigma_{\min}\sum_{i,j=1}^n A_{ij} = \sigma_{\min} \cdot \tr L \nonumber\\
&= \sigma_{\min} \sum_{i=2}^n \lambda_i
\ge \frac{\alpha_1}{\real\lambda_2} \sum_{i=2}^n \real\lambda_i \nonumber \\
&\ge \frac{\alpha_1}{\real\lambda_2} \cdot (n-1) \cdot \real\lambda_2 
= \alpha_1 (n-1).\label{eqn:bound2}\qed
\end{align}
\end{pf}
 
In the special cases where all the Laplacian eigenvalues are real, $S$ and $C$ can be computed explicitly.  In particular, $S$ is proportional to an eigenvalue ratio~\cite{barahona2002}.  Such situation occurs for example for the undirected networks, for which the Laplacian matrix is symmetric.  This result is expressed in the following theorem.
\begin{thm}\label{thm:real}
Suppose that the Laplacian eigenvalues are all real and ordered as $0 = \lambda_1 < \lambda_2 \le \cdots \le \lambda_n$.  
For any $\sr$ satisfying {\rm (A-I)} and {\rm (A-II)}, we have
\begin{equation}
S = \frac{\alpha_2}{\alpha_1} \cdot \frac{\lambda_2}{\lambda_n}, \quad
C = \frac{\alpha_1}{\lambda_2} \sum_{i=2}^n \lambda_i.
\end{equation}
\end{thm}
\begin{pf}
Suppose $\sr$ satisfies {\rm (A-I)} and {\rm (A-II)}.
Let $\sr_{\rm real}$ be the intersection of $\sr$ and the real axis.  
By combining (A-I) with the fact that $\sr$ is symmetric about the real axis, we see that $\sr_{\rm real}$ is an interval with endpoints at $\alpha_1$ and $\alpha_2$.
We can write
\begin{equation}
I_{\rm sync} = \bigcap_{i=2}^n I_{\rm sync}^{(i)}, \quad\text{where }
I_{\rm sync}^{(i)} =  \{ \sigma \,|\, \sigma\lambda_i \in \sr \}.
\end{equation}
Since $\lambda_i$ is real, we have $\sigma\lambda_i \in \sr$ if and only if $\sigma\lambda_i$ is in the interval $\sr_{\rm real}$, and hence $I_{\rm sync}^{(i)}$ is an interval whose endpoints are at $\alpha_1/\lambda_i$ and $\alpha_2/\lambda_i$.  Taking into account all $i=2,\ldots,n$, this means that $I_{\rm sync}$ is an interval with endpoints at $\sigma_{\min} =\alpha_1/\lambda_2$ and $\sigma_{\max} =\alpha_2/\lambda_n$. 
Thus,
\begin{equation}
S = \frac{\sigma_{\max}}{\sigma_{\min}} 
= \frac{\alpha_2}{\alpha_1} \cdot \frac{\lambda_2}{\lambda_n}.
\end{equation}
We also have
\begin{equation}
C = \sigma_{\min}\sum_{i,j=1}^n A_{ij} = \frac{\alpha_1}{\lambda_2} \cdot \tr L
= \frac{\alpha_1}{\lambda_2} \sum_{i=2}^n \lambda_i.\qed
\end{equation}
\end{pf}

A surprising consequence of this theorem is that a simple condition on the eigenvalues suffices to guarantee both the maximum synchronizability and the minimum synchronization cost for any dynamical component $(\mathbf{F},\mathbf{H},\mathbf{s})$ with $\sr$ satisfying (A-I) and (A-II):

\begin{cor}\label{cor:optimal}
Suppose that the Laplacian eigenvalues of a network satisfy,
\begin{equation}
0=\lambda_1 < \lambda_2=\cdots=\lambda_n.
\label{eqn:optimal}
\end{equation} 
Then, $S$ and $C$ achieve their maximum and minimum values, respectively, {\it i.e.},
\begin{equation}
S = \frac{\alpha_2}{\alpha_1}, \quad
C = \alpha_1(n-1),
\end{equation}
for any $(\mathbf{F},\mathbf{H},\mathbf{s})$ with $\sr$ satisfying {\rm (A-I)} and {\rm (A-II)}.
\end{cor}
\begin{pf}
Condition~\eqref{eqn:optimal} implies that the eigenvalues are all real, so Theorem~\ref{thm:real} applies, and we have
\begin{equation}
S = \frac{\alpha_2}{\alpha_1} \cdot \frac{\lambda_2}{\lambda_n}
= \frac{\alpha_2}{\alpha_1},
\end{equation}
and
\begin{equation}
C = \frac{\alpha_1}{\lambda_2} \sum_{i=2}^n \lambda_i = \alpha_1 (n-1).\qed
\end{equation}
\end{pf}
Corollary~\ref{cor:optimal} shows that any network satisfying~\eqref{eqn:optimal} is a {\em simultaneous} solution of the two optimization problems under consideration.  However, even more surprising is the fact that those that satisfy~\eqref{eqn:optimal} are actually the {\em only} solutions.   In other words, the two classes of optimal networks---those with the maximum synchronizability and those with the minimum synchronization cost---actually coincide, and both can be completely characterized by condition~\eqref{eqn:optimal}.
To state this in a precise but convenient form, let us define the following terminology.  We say that a network given by $A$ has {\em the maximum synchronizability} if $S = \alpha_2/\alpha_1$ for any $(\mathbf{F},\mathbf{H},\mathbf{s})$ with $\sr$ satisfying (A-I) and (A-II).  Similarly, we say that a network given by $A$ has {\em the minimum synchronization cost} if $C = \alpha_1(n-1)$ for any $(\mathbf{F},\mathbf{H},\mathbf{s})$ with $\sr$ satisfying (A-I) and (A-II).  Thus, the optimality is a property that depends solely on the network structure and not on the dynamical component of the system. 

\begin{thm}\label{thm:optimal}
The following statements are equivalent:
\renewcommand{\labelenumi}{\rm (\roman{enumi})}
\begin{enumerate}
\item A network has the maximum synchronizability.
\item A network has the minimum synchronization cost.
\item The Laplacian eigenvalues of a network satisfy condition~\eqref{eqn:optimal}.
\end{enumerate}
\end{thm}
\begin{pf}
(iii) $\Rightarrow$ (i) and (iii) $\Rightarrow$ (ii) are precisely what Corollary~\ref{cor:optimal} states, so we are left with proving (i) $\Rightarrow$ (iii) and (ii) $\Rightarrow$ (iii).  We do this by showing their contrapositives, {\it i.e.}, that if (iii) does not hold, then neither (i) nor (ii) hold. If (iii) does not hold, we either have $\real\lambda_2 < \real\lambda_n$ or $\imag\lambda_k \neq 0$ for some $k$, where $2 \le k \le n$.  If $\real\lambda_2 < \real\lambda_n$, then, by \eqref{eqn:bound} and \eqref{eqn:bound2}, we have
\begin{equation}
S  \le \frac{\alpha_2}{\alpha_1} \cdot \frac{\real\lambda_2}{\real\lambda_n}
< \frac{\alpha_2}{\alpha_1},
\end{equation}
and
\begin{equation}
\begin{split}
C &\ge \frac{\alpha_1}{\real\lambda_2} \sum_{i=2}^n \real\lambda_i \\
&\ge \alpha_1 \left[ (n-2) +\frac{\real\lambda_n}{\real\lambda_2} \right]\\
&> \alpha_1 (n-1),
\end{split}
\end{equation}
and hence neither (i) nor (ii) hold.  If $\real\lambda_2=\cdots=\real\lambda_n$ and $\imag\lambda_k \neq 0$ for some $k$, then there exist systems with $\sr$ satisfying (A-I) and (A-II) (in fact, most systems; see Fig.~\ref{fig:proof}) such that $\{ \real(\sigma\lambda_k) \,|\, \sigma\lambda_k \in \sr \}$ is an interval with endpoints at $\alpha'_1$ and $\alpha'_2$, where $\alpha_1 < \alpha'_1<\alpha'_2 < \alpha_2$, implying
\begin{equation}
\frac{\alpha'_2}{\alpha'_1} < \frac{\alpha_2}{\alpha_1}.
\end{equation}
\begin{figure}[t]
\begin{center}
\epsfig{file=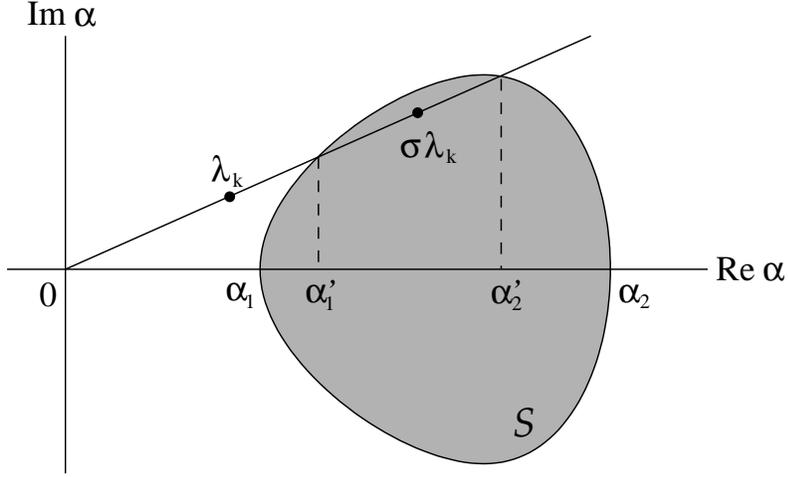,width=0.75\textwidth}
\caption{A situation leading to suboptimal $S$ and $C$ in the proof of Theorem~\ref{thm:optimal}.}
\label{fig:proof}
\end{center}
\end{figure}
Here we used the fact that $\sr_{\rm real}$ is an interval with endpoints at $\alpha_1$ and $\alpha_2$, as in the proof of Corollary~\ref{cor:optimal}.
Again, by using the same argument as in the proof of Theorem~\ref{thm:bounds} with $\alpha_1$ and $\alpha_2$ replaced with $\alpha'_1$ and $\alpha'_2$, we have
\begin{equation}
S \le  \frac{\alpha'_2}{\alpha'_1} <  \frac{\alpha_2}{\alpha_1},
\end{equation}
and 
\begin{equation}
C \ge \alpha'_1 (n-1) > \alpha_1 (n-1),
\end{equation}
so neither (i) nor (ii) hold.  We have shown (i) $\Leftrightarrow$ (iii) and (ii) $\Leftrightarrow$ (iii), which prove the equivalence of the three statements. \qed
\end{pf}
The conclusions about the synchronization cost $C$ in Theorem~\ref{thm:bounds}, Theorem~\ref{thm:real}, Corollary~\ref{cor:optimal}, and Theorem~\ref{thm:optimal} remain valid if the assumptions on the stability region are weakened, with straightforward modification to the proofs.  
The bound on $C$ in Theorem~\ref{thm:bounds} and the identity for $C$ in Theorem~\ref{thm:real} are valid without assuming (A-II).  The conclusion about $C$ in Corollary~\ref{cor:optimal} remains valid if (A-I) and (A-II) are replaced with the condition that $\alpha_1$ (as a point in the complex plane) is contained in $\sr_{\rm real}$.
The latter condition is also sufficient for the equivalence between statements (ii) and (iii) in Theorem~\ref{thm:optimal}.

\begin{figure}[t]
\begin{center}
\epsfig{figure=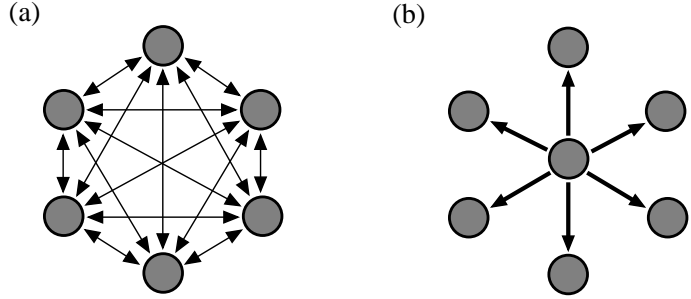,width=0.65\textwidth}
\caption{Simple examples of optimal networks.  (a) The global coupling configuration with uniform weight $\lambda/n$ on every link, where $n=6$ and each double arrow represents two links, one in each direction.  (b) The outward oriented star configuration with uniform weight $\lambda$ on every link, where each arrow represents a single link.}
\label{fig:optimal1}
\end{center}
\end{figure}

In view of the equivalence in Theorem~\ref{thm:optimal} and for convenience, we will use the following terminology in the rest of the paper.
\begin{defn}
We say that a network given by $L$ is \textbf{optimal} if it satisfies condition~\eqref{eqn:optimal}.
\end{defn}
Thus, the class of optimal networks is the set of simultaneous solutions to the two optimization problems.
The uniform global coupling, in which all oscillators are connected to all the other oscillators with weight $\lambda/n$ on all links, is perhaps the simplest example of an optimal network [Fig.~\ref{fig:optimal1}(a)].  The non-zero Laplacian eigenvalues in this case are $\lambda_2 = \dotsb = \lambda_n = \lambda$, and therefore this network satisfies condition~\eqref{eqn:optimal}.  Another simple example is the {\em outward oriented star}, defined here as having a single node connected to all the other nodes with uniform weight $\lambda$ on all of these links [Fig.~\ref{fig:optimal1}(b)].  The Laplacian eigenvalues are also $\lambda_2 = \dotsb = \lambda_n = \lambda$ for this configuration.
However, there are many more networks that are optimal, as we will see in Section~\ref{sec:hierarchy}.

\section{Diagonalizability of Optimal Structures\label{sec:non-diag}}

\begin{figure}[t]
\begin{center}
\epsfig{figure=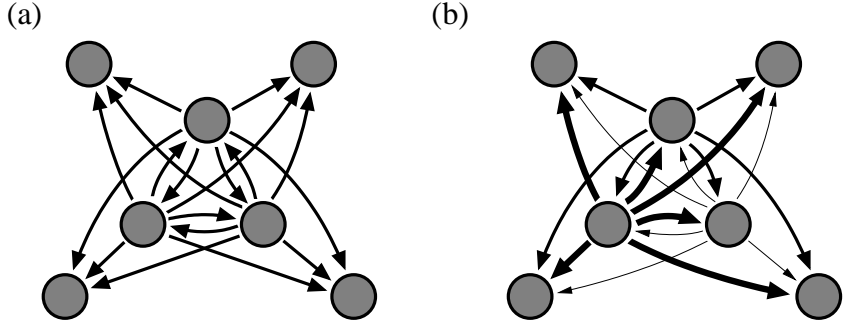,width=0.8\textwidth}
\caption{Examples of optimal networks that are diagonalizable.  Each arrow represents a single directed link.  Thick, medium, and thin arrows have weight $\lambda/2$, $\lambda/3$, and $\lambda/6$, respectively.  For each network, all nonzero eigenvalues are $\lambda$.}
\label{fig:optimal2}
\end{center}
\end{figure}

We now study the diagonalizability of the optimal networks defined by~\eqref{eqn:optimal}.  Here we say that a network is diagonalizable if the corresponding Laplacian matrix is diagonalizable.  Otherwise, the network is called non-diagonalizable.  We will show that all networks that are diagonalizable must satisfy a special structural condition.
This has a rather surprising consequence that {\em most optimal networks are non-diagonalizable}.  
This means that the extension of the master stability analysis in Section~\ref{sec:extension} was indeed necessary for a proper treatment of the optimization problems.  The following theorem gives a complete characterization of the optimal networks that are diagonalizable.
\begin{thm}\label{thm:non-diag}
The following two statements about a given network are equivalent:
\begin{enumerate}
\item[\rm (i)] The network is optimal and diagonalizable.
\item[\rm (ii)] The oscillators can be divided into two groups: those with uniform positive output to all the other oscillators and those with no output at all.  
In addition, there is at least one oscillator in the first group.
\end{enumerate}
\end{thm}
\begin{pf}
Suppose that a network satisfies condition (i), {\it i.e.}, $L$ is optimal, so that $\lambda=\lambda_2=\cdots=\lambda_n > 0$, and $L$ is diagonalizable.  Then, the eigenspace associated with the eigenvalue $\lambda$ has the maximum possible dimension of $n-1$, and so does the solution space of the eigenvalue equation $Lx=\lambda x$, which can also be written as $(L-\lambda I)x = 0$.  This implies that all rows of the matrix $L-\lambda I$ must be constant multiples of the first row, so the matrix must be of the form
\begin{equation}\label{eqn:L-lambda}
L-\lambda I = \begin{pmatrix}
a_1 & a_2 & \cdots & a_n\\
c_2 a_1 & c_2 a_2 & \cdots & c_2 a_n\\
\vdots & \vdots & & \vdots\\
c_n a_1 & c_n a_2 & \cdots & c_n a_n
\end{pmatrix}.
\end{equation}
Then, the Laplacian matrix itself takes the form
\begin{equation}
L = \begin{pmatrix}
a_1 + \lambda & a_2 & \cdots & a_n\\
c_2 a_1 & c_2 a_2 + \lambda & \cdots & c_2 a_n\\
\vdots & \vdots & & \vdots\\
c_n a_1 & c_n a_2 & \cdots & c_n a_n + \lambda
\end{pmatrix}.
\end{equation}
However, from the property of a Laplacian matrix that $\sum_j L_{ij} = 0$, it follows that $c_2=\cdots =c_n = 1$, and hence
\begin{equation}\label{eqn:L}
L = \begin{pmatrix}
a_1 + \lambda & a_2 & \cdots & a_n\\
a_1 & a_2 + \lambda & \cdots & a_n\\
\vdots & \vdots & & \vdots\\
a_1 & a_2 & \cdots & a_n + \lambda
\end{pmatrix}.
\end{equation}
The off-diagonal entries in the $j$th column represent the strength of the output from the $j$th oscillator to the other oscillators.  Hence, this form of $L$ implies that each oscillator $j$ either has the same positive output to all the oscillators ($a_j \neq 0$) or it has no output at all ($a_j = 0$).  In addition, there must be at least one oscillator that has positive output, since otherwise the network is completely disconnected, and it is impossible for it to be optimal.

Now suppose that the network satisfies condition (ii).  
Then, since the strength $A_{ij}$ of the connection from node $j$ to node $i \neq j$ depends only on $j$, the adjacency matrix $A$ must have the form
\begin{equation}
A = \begin{pmatrix}
\ 0\ & \ b_2\ &\ \cdots\ &\  b_n\\
b_1 & 0 & \cdots & b_n\\
\vdots & \vdots & & \vdots\\
b_1 & b_2 & \cdots & 0
\end{pmatrix},
\end{equation}
where $b_i \ge 0$.  Using $\sum_j L_{ij} = 0$, we can show that the Laplacian matrix $L$ must be in the form
\begin{equation}
L = \begin{pmatrix}
\lambda - b_1 & -b_2 & \cdots & -b_n\\
-b_1 & \lambda - b_2 & \cdots & -b_n\\
\vdots & \vdots & & \vdots\\
-b_1 & -b_2 & \cdots & \lambda - b_n
\end{pmatrix},
\end{equation}
where $\lambda = \sum_i b_i > 0$.  From this, it is straightforward to show that the eigenvalues of $L$ are $0 = \lambda_1 < \lambda_2 = \dotsb = \lambda_n = \lambda$, and the eigenspace of $\lambda$ has dimension $n-1$.  Thus, the network is optimal and diagonalizable.\qed
\end{pf}
The uniform global coupling and the outward oriented star configurations, shown in Fig.~\ref{fig:optimal1}, are examples of optimal networks that are diagonalizable.  Figure~\ref{fig:optimal2} shows two more such examples.  For both examples, the Laplacian eigenvalues $\lambda_2,\dotsc,\lambda_7$ are all equal to $\lambda$ and the corresponding eigenspace has the full dimension of 6.

In fact, condition (ii) in Theorem~\ref{thm:non-diag} provides us with an explicit procedure to construct all optimal networks that are diagonalizable.  
Start with a set of $n$ nodes with no links and choose nonnegative numbers $b_1, b_2, \dotsc, b_n$, not all zero. 
For each $i$, make directed links from node $i$ to the $n-1$ other nodes with weight $b_i$ on each link.
Regardless of the choice of $b_i$'s, the resulting network is guaranteed to satisfy condition (ii), and different choices of $b_i$'s are guaranteed to generate all such networks.
In the example of Fig.~\ref{fig:optimal2}(a), the three nodes in the middle have links to all the other nodes with weight $b_i = \lambda/3$, while the remaining nodes have no outgoing links ($b_i = 0$).  In Fig.~\ref{fig:optimal2}(b), the three middle nodes have outgoing links with different weights ($b_i = \lambda/2, \lambda/3, \lambda/6$) that add up to $\lambda$.
In this class of networks, the $(n-1)$-degenerate eigenvalue is always equal to $\sum_i b_i$.

The theorem implies that for any optimal network that is diagonalizable, there is at least one oscillator with uniform positive output to all the other oscillators in the network.  This, however, is unlikely to occur in a large realistic complex networks, implying that if synchronization is important for such networks, they are likely to be non-diagonalizable, or at least close to being non-diagonalizable.

\section{Optimality of Hierarchical Network Structures\label{sec:hierarchy}}

Here we present another subclass of optimal networks, characterized by three structural conditions that are more intuitive than~\eqref{eqn:optimal}. The first condition ensures connectedness of the network, the second ensures well-defined hierarchy of nodes, and the third ensures uniformity of total input strength in each node.  To conveniently state the first condition, we define an {\em oriented spanning tree} to be a directed subnetwork that is a tree and spans the entire set of nodes, with the links oriented in the direction from the root to the branches of the tree.  Thus, the existence of an oriented spanning tree embedded in a network is equivalent to the existence of a node from which all other nodes can be reached by following the directed links.
\begin{thm}\label{thm:optimal2}
Suppose that a network satisfies the following three conditions:
\begin{enumerate}
\item[\rm (i)] It embeds an oriented spanning tree.
\item[\rm (ii)] It has no feedback loops.
\item[\rm (iii)] For all nodes that receive positive input, the sum of input strength $\sum_{j\neq i} A_{ij} = L_{ii} $ is equal to a constant $\lambda$.
\end{enumerate}
Then, the network is optimal and the $(n-1)$-degenerate eigenvalue is equal to $\lambda$. 
\end{thm}
\begin{pf}
Suppose that a network satisfies the given conditions~(i)--(iii).  By starting from an arbitrary node and traversing nodes following links in the {\em reverse} direction, we must eventually either return to a node already visited, thus creating a feedback loop, or arrive at a node without any input.  By condition~(ii) we cannot have any feedback loops, so we must arrive at a node that receives no input.   Such a node can only be the unique node at the root of the oriented spanning tree that is guaranteed to exist by condition (i), since any other nodes in the tree must be reachable from the root.  

Now let us assign the index 1 to the unique node without input.  
Consider the network obtained by removing the node 1 and any links from it.  Applying the same argument as above to this subnetwork (but now only the existence part), we see that there is at least one node without input within the subnetwork.  The only input to such a node is from node 1.  Let $n_2$ be the number of such nodes.  We arbitrarily index them $2, 3, \ldots, n_2+1$.  Let us then consider the network obtained by removing these $n_2$ additional nodes, together with all associated links.  Applying the same argument again, we see that there is at least one node whose only input is from nodes $1, 2, \ldots, n_2+1$, and we index them $n_2+2, n_2+3, \ldots, n_2+n_3+1$.  Repeating this argument until we assign indices to all the nodes, we obtain an indexing in which all the links are from a node with smaller index to a node with larger index.  This means that the Laplacian matrix $L$ of the network using these indices is a lower triangular matrix, and hence the diagonal elements are its eigenvalues.  Since the diagonal elements of $L$ are precisely the total input strength of the nodes, the first one is 0, corresponding to the unique node without input, and all the others are $\lambda$, which follows from condition (iii).  Hence the network satisfies the condition~\eqref{eqn:optimal} and therefore is optimal, by Theorem~\ref{thm:optimal}.\qed
\end{pf}

Note that condition~(i) of Theorem~\ref{thm:optimal2} is equivalent to $\real\lambda_2 > 0$, which follows immediately from a recent result in Ref.~\cite{Wu2005}, and this generalizes the notion of connectedness to directed networks.  Condition~(i) is necessary for a network to satisfy the stability condition~\eqref{eqn:stability}.  In other words, the network must be connected in this sense to make sure that it is at least compatible with the possibility of stable complete synchronization.

\begin{figure}[t]
\begin{center}
\epsfig{figure=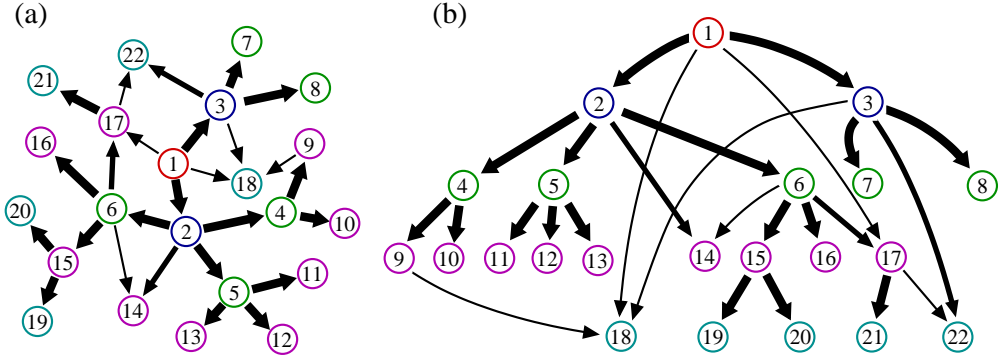,width=0.95\textwidth}
\caption{(Color online) (a) Example of optimal network with hierarchical structure.  Thick, medium, and thin arrows have weight $\lambda$,  $2\lambda/3$, and $\lambda/3$, respectively, where the sum of input strengths in each node is normalized to $\lambda$.  The nodes are numbered according to the ranking and colored by the levels in the hierarchy.  (b) The nodes in (a) are rearranged to make the hierarchical levels more clearly visible. \label{fig:hierarchy}}
\end{center}
\end{figure}

In Fig.~\ref{fig:hierarchy}, we show an example of a network satisfying conditions (i)--(iii) of Theorem~\ref{thm:optimal2}.  
The method of assigning indices to the nodes described in the proof above defines a ranking of the nodes in such a network.  In this ranking, links exist only from a node of higher rank to a node of lower rank.  In addition, the method defines a hierarchical structure with multiple levels.  The top level contains the unique node without input.  The second level consists of $n_2$ nodes that receive input only from the top level.  The third level consists of $n_3$ nodes that receive input only from the top two levels.  The rest of the hierarchical structure is defined similarly, so that links exist only from a higher level to a lower level.   These hierarchical levels are indicated by different colors in Fig.~\ref{fig:hierarchy}, and in panel (b), the nodes are rearranged to make the levels more clearly visible.  

The flow of information about the dynamical state of the oscillators in this hierarchical structure is unidirectional; it flows only from the top down to the bottom.  With this picture in mind, the reason for the optimality of such a network can be understood intuitively as follows.  The top node in the hierarchy receives no input and acts as the ``master'' oscillator that dominates the network dynamics. 
If the coupling strength $\sigma$ is chosen so that 
$\Lambda(\sigma\lambda) < 0$, where $\lambda>0$ is the $(n-1)$-degenerate eigenvalue, then the oscillators in the second level, which receive input only from the master, will synchronize themselves with the master.  
An oscillator in the third level, which receive input only from those in the top two levels, must also synchronize, since normalization of the total input strength makes the equation effectively look as if it were receiving input from a single oscillator synchronized with the master.  Repeating the same argument for the rest of the hierarchical levels in the network, we see that all oscillators
must eventually synchronize.  Thus, conditions (i)--(iii) guarantee stable synchronization in the entire range of $\sigma$ such that $\Lambda(\sigma\lambda) < 0$.  This makes perfect sense because the stability condition~\eqref{eqn:stability} becomes $\Lambda(\sigma\lambda) < 0$ when the optimality condition~\eqref{eqn:optimal} is satisfied.  

Notice that this argument is very similar to the argument in Section~\ref{sec:extension} that was used to derive the stability condition for non-diagonalizable cases. 
This suggests that the networks satisfying conditions (i)--(iii) may also suffer from long transient before converging to the synchronized state.  For these networks, the number of levels in the hierarchy is strongly related to the length of the transient.  In addition, the similarity suggests that most of these hierarchical networks in Theorem~\ref{thm:optimal2} are non-diagonalizable.  In fact, the only exception is when there are only two levels in the hierarchy, with one top node connected to all the other nodes with uniform weights, leading to the outward oriented star configuration.  
\begin{thm}
Let $L$ represent a network satisfying {\rm (i)--(iii)} in Theorem~\ref{thm:optimal2}.  Then, $L$ is diagonalizable if and only if it is the outward oriented star.
\end{thm}
\begin{pf}
Suppose $L$ is diagonalizable.  Then, by Theorem~\ref{thm:non-diag}, for each $i$, oscillator $i$ either has the same nonzero output to all the other oscillators or has no output at all.  From the argument in the proof of Theorem~\ref{thm:optimal2}, there is a unique oscillator without any input.  This oscillator must have uniform output to all the other oscillators, since it would be isolated otherwise.  Any of the other oscillators must have no output at all, since otherwise it would have output to all the other oscillators, including the first one, which leads to a feedback loop.  Thus, we have have the outward oriented star configuration, and the weights on the links are uniform because of the condition (iii) in Theorem~\ref{thm:optimal2}.

If the network is the outward oriented star, then it is clear that each oscillator either has the same nonzero output to all the other oscillators or has no output at all.  Therefore, by Theorem~\ref{thm:non-diag}, it must be diagonalizable. \qed
\end{pf}

This result is intimately related to the structure of branches in the underlying spanning trees.  In an oriented spanning tree, it can be shown that the number of independent eigenvectors associated with an $(n-1)$-degenerate eigenvalue $\lambda >0$ is equal to the number of branches in the tree.

\section{Optimization with Topological Constraints}

Let us now consider optimization problems with topological constraints.  Suppose that the oscillators are constrained to interact only within a given network topology represented by a symmetric matrix $A_0$ defined by
\begin{equation}
(A_0)_{ij} = \begin{cases}
1, & \text{if distinct oscillators $i$ and $j$ are allowed to interact,}\\
0, & \text{otherwise}.
\end{cases}
\end{equation}
Note that $A_0$ represents an {\em undirected} network of interaction topology.
To make the system compatible with synchronization, we assume that this network is connected, {\it i.e.}, there is a path between any two nodes.  The problem is to choose an assignment of weights and directions for the links in this network, so that the resulting network is optimal.  Let $W_{ij} \ge 0$ be the weight assigned to the directed link from $j$ to $i$.  With this assignment, we obtain a network with adjacency matrix $A$ given by $A_{ij} = (A_0)_{ij}W_{ij}$.  Then, the constrained optimization problems can be formulated as follows.  Given a connected network topology $A_0$ of allowed interactions, for which assignment $W$ does the resulting network have the maximum synchronizability and/or the minimum synchronization cost?

Remarkably, there always exists an assignment that achieves the optimality defined by~\eqref{eqn:optimal} for any constraining topology $A_0$ that is connected.  We can explicitly construct solutions using Theorem~\ref{thm:optimal2}, together with the fact that we can always find an oriented spanning tree within the topology $A_0$.   
Indeed, if we properly assign directions to links along such a tree, then conditions (i) and (ii) are clearly satisfied. 
The fact that properly weighted trees can have identical nonzero eigenvalues has been noted before, without considering their non-diagonalizability~\cite{sb}.

One way to explicitly construct an oriented spanning tree is the well-known procedure called the breadth-first search.  The procedure also determines a ranking of the nodes and the hierarchical levels.  First, we choose an arbitrary node as the top rank node which forms the top hierarchical level by itself.  Then, we find all nodes that can be reached from the first node, make connections to them, and rank them arbitrarily following the first node.  These nodes form the second level in the hierarchy.  The third level consists of the nodes that are not yet discovered but can be reached by following two links, and again we rank them arbitrarily following the nodes already discovered.  We make connections to these nodes, making sure to choose only one connection to each node.  We repeat this until we discover all the nodes in the network, and the resulting directed network is an oriented spanning tree.
This procedure can produce at least $n$ distinct oriented spanning trees, one for each choice of the root node, but in general there are many others, and many of them cannot be found by this procedure.  Indeed, from the Matrix-Tree Theorem it follows that the number of all such oriented trees is $\Pi_{i=2}^{n}\mu_i$,  where $\mu_2,\ldots,\mu_n$ are the
nonzero Laplacian eigenvalues of the underlying {\it undirected} network defined by matrix $A_0$.
For a globally connected network, for example, the number of oriented spanning 
tree is $n^{n-1}$, which is a huge number even for relatively small networks. 

\begin{figure}[t]
\begin{center}
\epsfig{figure=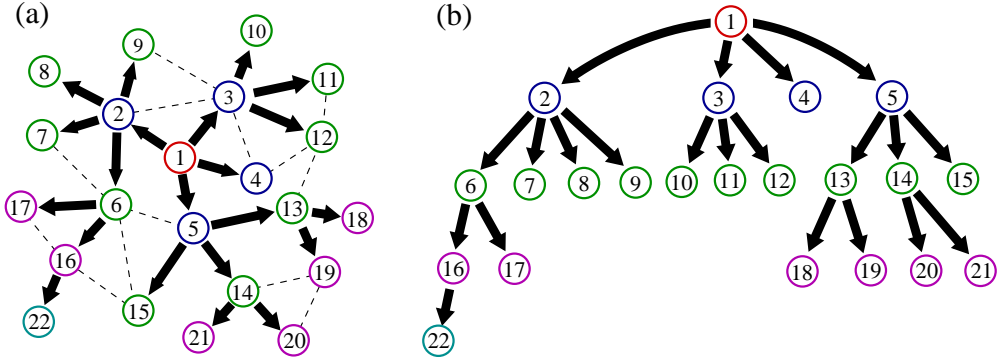,width=0.95\textwidth}
\caption{(Color online) (a) Example of weight and direction assignment within a given interaction topology based on an oriented spanning tree constructed by the breadth-first search.  Arrows indicate directed links with nonzero weight, while dashed lines are links with zero weight.  (b) The nodes in (a) are rearranged to make the hierarchical levels more clearly visible. \label{fig:ost}}
\end{center}
\end{figure}

To ensure that condition (iii) is also satisfied, we assign the same weight to all the links in the oriented spanning tree.  The resulting network is guaranteed by Theorem~\ref{thm:optimal2} to have the maximum synchronizability and the minimum synchronization cost (see Fig.~\ref{fig:ost} for an example).  
Therefore, for a given connected topology of possible interactions, there are at least as many solutions to the constrained optimization problems as the number of oriented spanning trees.  
However, there are certainly many more ways to assign weights and directions so as to satisfy the conditions in Theorem~\ref{thm:optimal2}.   
For example, any addition of directed links to an oriented spanning tree, that is allowed by the topology of possible interactions and that does not create
loops, leads to an optimal network after normalization of inputs.
It is rather remarkable that these solutions allow the network with an arbitrary topological constraint to achieve the same synchronization efficiency---from both synchronizability and cost viewpoints---as the global coupling configuration.

An interesting property of this construction is that the choice of the ``master''
oscillator does not matter in achieving optimality. Despite the intuition
that the nodes with the largest number of links in the given topology are the most
natural choices for the master, the above construction shows that {\it any} node
can be the master. Moreover, the direction of each link in an optimal network
is not necessarily related to the properties of the nodes it connects.

Because the weight assignments based on oriented spanning trees explore only some of the many potential links, the optimality of the resulting networks can be interpreted as a synchronization version of the paradox of Braess for traffic flow~\cite{Braess1968,Irvine:1993}, in which removing links leads counter-intuitively to improved performance of the network.  
It is interesting to notice that similar directed networks without
feedback loops also emerge as gradient networks~\cite{toroczkai2004} and in the study of
transport processes on complex networks~\cite{Durand}.

An immediate consequence of Theorem~\ref{thm:non-diag} is that all solutions of the constrained optimization problem are non-diagonalizable, unless some oscillator is allowed to interact with all the other oscillators. 
\begin{cor}
Suppose that the topology $A_0$ of interactions allows no oscillator to interact with all the other oscillators.  If an assignment of weights and directions leads to an optimal network (which is always possible as long as $A_0$ is connected), then the resulting network is non-diagonalizable.
\end{cor}
\begin{pf}
Suppose that the optimal network obtained by the assignment of weights and directions is diagonalizable. Theorem~\ref{thm:non-diag} implies that there must be at least one oscillator having nonzero output to all the other oscillators.  Therefore, under the assumption of the Corollary, the network must be non-diagonalizable. \qed
\end{pf}
Once again, this shows that the extension of the master stability framework was indeed necessary to treat the problem.  The global coupling topology and star topology are among the exceptional cases where an oscillator can communicate with all the other oscillators, but such a situation is uncommon in a large complex network. 

The oriented spanning trees form a subclass of all optimal networks constrained under the given topology of interactions.  One property that sets the oriented spanning trees apart from others in the class of optimal networks is that in addition to maximizing synchronizability and minimizing the synchronization cost, it also minimizes the number of links used ({\it i.e.}, the number of links with nonzero weight).  This is because the number of links in an oriented spanning tree, which is always $n-1$, is the minimum number of links required to span all $n$ nodes.

The oriented spanning trees can be used as a basis for optimization under even stricter constraints. Suppose that, in addition to constraining the topology of interactions, we have constraints on the directions of allowed links.  In other words, we consider optimization among all subnetworks of a given directed network represented by
\begin{equation}
(A_0)_{ij} = \begin{cases}
1, & \text{if oscillator $i$ may receive connection from oscillator $j \neq i$,}\\
0, & \text{otherwise}.
\end{cases}
\end{equation}
Note that in this case, $A_0$ is not necessarily symmetric.  As long as the directed network given by $A_0$ is connected in the sense that it embeds an oriented spanning tree, explicit construction of an optimal subnetwork is possible.  
These optimal subnetworks include the embedded oriented spanning tree itself and any other subnetwork satisfying the conditions in Theorem~\ref{thm:optimal2}.

Interestingly enough, despite all the optimal properties that stem from the properties of oriented spanning trees, {\it undirected} tree topology was found to be among the most difficult to synchronize~\cite{Yook2005}.  
Moreover, the out-degree distribution of these optimal networks can be highly heterogeneous, in sharp contrast with the case of undirected networks~\cite{Nishikawa2003}.
These highlight the significance of directionality of the interactions in determining the synchronizability of networks.  The fact that directed networks may have advantage over undirected ones is consistent with the finding in~\cite{Motter2005b,Motter2005c,Motter2005,Zheng2000} that asymmetric coupling in networks of chaotic oscillators has positive effect on synchronization.

\section{Concluding Remarks}

In this work, we have considered the problem of maximizing the synchronizability and minimizing the synchronization cost of oscillator networks.   By extending the master stability formalism to the case of non-diagonalizable Laplacian matrices, we have shown that the solutions of the optimization problems can be completely characterized by the simple condition~\eqref{eqn:optimal} on the Laplacian eigenvalues.  We have also shown that the intuitive structural conditions (i)--(iii) in Theorem~\ref{thm:optimal2}, which facilitate unidirectional information flow and normalized input strength, guarantee optimality. In addition, we have considered optimization under topological constraints and shown that we can explicitly construct optimal networks using oriented spanning trees.  Furthermore, by a complete characterization of diagonalizable optimal networks, we have proved that most optimal networks are actually non-diagonalizable, which necessitates the extension of the master stability formalism.
Since spectral analyses are also relevant for other dynamical processes on networks (e.g., diffusion and other spreading processes), it would be interesting to see these results applied to the study of different network phenomena.

Structural properties of optimal networks, such as those given in our theorems for the diagonalizable and hierarchical networks, can serve as a general guideline for designing networks for synchronization.  For such applications, however, it is important to address the question of robustness.
What is the effect of structural perturbations on the optimality of these networks?
We expect that a perturbation theory can be used to show that small deviations from the optimal structures will induce only a small change in $S$ and $C$.  
Another type of robustness question is about the effect of dynamical perturbations:  how does disturbances introduced at one or more nodes in the synchronized state propagate over the network?  

\begin{figure}
\begin{center}
\epsfig{figure=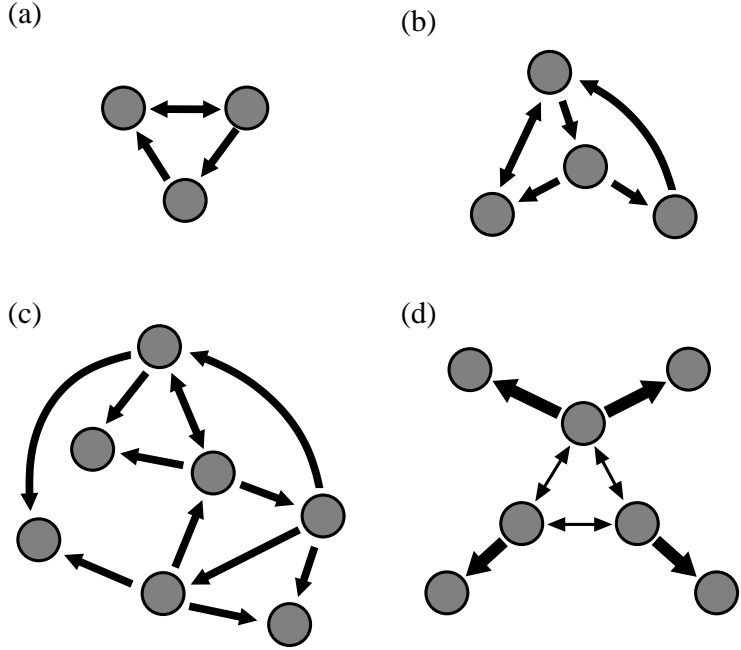,width=0.7\textwidth}
\caption{Examples of optimal networks that are neither diagonalizable nor hierarchical.  An arrow indicates a directed link, while a double arrow indicates two directed links.  Thick, medium, and thin arrows have weight $\lambda$, $\lambda/2$, and $\lambda/3$, respectively.  For each network, all nonzero eigenvalues are $\lambda$.}
\label{fig:optimal3}
\end{center}
\end{figure}
\begin{figure}
\begin{center}
\epsfig{figure=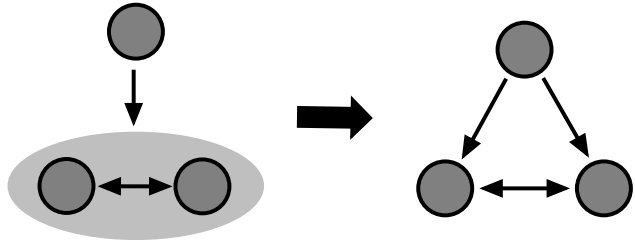,width=0.6\textwidth}
\caption{Example of a suboptimal network resulting from combining two optimal networks.  The lower-level node of a hierarchical network with two nodes is replaced by a global coupling network with two nodes.  We can show that no combinations of weights on the two downward links can make this network optimal.}
\label{fig:subopt}
\end{center}
\end{figure}

Within the class of all optimal networks, we have explicitly constructed a large number of important networks: those that are diagonalizable (Theorem~\ref{thm:non-diag}) and those that are hierarchical (Theorem~\ref{thm:optimal2}).  
However, there are many other optimal networks that fall into neither categories (see Fig.~\ref{fig:optimal3} for examples). 
Some of them can be constructed by combining two optimal networks.
For example, the network shown in Fig.~\ref{fig:optimal3}(d) is a combination of a hierarchical network and a diagonalizable network, constructed by replacing the root node in the 5-node outward oriented star (hierarchical) with the 3-node global coupling configuration (diagonalizable).  
However, the other examples in Fig.~\ref{fig:optimal3} cannot be constructed in a similar fashion.  There are also cases in which this kind of construction does not lead to an optimal network (see Fig.~\ref{fig:subopt} for an example).  
The explicit construction of all optimal networks is an important open problem.

The complete characterization of the entire class of optimal networks is expected to have a profound implication on the widely assumed hypothesis that synchronizability plays an important role in the evolution of many real-world complex networks.  If the signatures typical of optimal networks are found to be absent in these networks, then the hypothesis may be questionable; synchronization may be less significant than other competing factors such as robustness, or else, the oscillator network model does not describe the essential features of the system.  
Hierarchical structures have been suggested to play a significant role in the network of motor neurons of {\em Aplysia}~\cite{Jing2005} and in the mechanism of invariant visual representation
in the human brain~\cite{Quiroga2005}.   
Hierarchical structures are also common in many human organizations, perhaps because they can better facilitate coordinated activities. It is important to examine the existing real data as a first step toward testing this and possibly other hypotheses about the evolution of complex networks.


\begin{thebibliography}{99}

\bibitem{Zhou:2006b}
C.~S.~Zhou, A.~E. Motter, and J.~Kurths,
Phys. Rev. Lett. {\bf 96}, 034101 (2006).

\bibitem{Li:2006}
C.~Li, L.~Chen, and K.~Aihara,
Physical Biology {\bf 3}, 37 (2006).

\bibitem{Stilwell:2006}
D.~J. Stilwell, E.~M. Bollt, and D.~G. Roberson, 
SIAM J. Appl. Dyn. Syst. {\bf 5}, 140 (2006).

\bibitem{Arenas:2006}
A.~Arenas, A.~D\'iaz-Guilera, and C.~J. P\'erez-Vicente,
Phys. Rev. Lett. {\bf 96}, 114102 (2006).

\bibitem{Wiley:2006}
D.~A. Wiley, S.~H. Strogatz, and M.~Girvan,
Chaos {\bf 16}, 015103 (2006).

\bibitem{Park2006}
S. M. Park and B. J. Kim, Phys. Rev. E {\bf 74}, 026114 (2006).

\bibitem{Boccaletti2006}
S. Boccaletti, D.-U. Hwang, M. Chavez, A. Amann, J. Kurths, and L. M.
Pecora, Phys. Rev. E {\bf 74}, 016102 (2006).

\bibitem{Hasegawa2006}
H. Hasegawa, Phys. Rev. E {\bf 72}, 056139 (2005).

\bibitem{Donetti:2005}
L.~Donetti, P.~I. Hurtado, and M.~A. Mu{\~n}oz, 
Phys. Rev. Lett. {\bf 95}, 188701 (2005).

\bibitem{Morelli2005}
L.~G. Morelli, H. Cerdeira, and D.~H. Zanette, Eur. Phys. J. B {\bf 43}, 243 (2005).

\bibitem{Restrepo2005}
J.~G. Restrepo, E. Ott, and B. R. Hunt,  Phys. Rev. E {\bf 71}, 036151 (2005).

\bibitem{Oh:2005}
E.~Oh, K.~Rho, H.~Hong, and B.~Kahng,
Phys. Rev. E {\bf 72}, 047101(2005).

\bibitem{Grinstein:2005}
G.~Grinstein and R.~Linsker,
Proc. Nat. Acad. Sci. {\bf 102}, 9948 (2005).

\bibitem{Wu:2005}
C.~W. Wu,
Nonlinearity {\bf 18}, 1057 (2005).

\bibitem{Masoller:2005}
C.~Masoller and A.~C. Mart\'i,
Phys. Rev. Lett. {\bf 94}, 134102 (2005).

\bibitem{Jalan:2005}
S.~Jalan, R.~E. Amritkar, and C.-K. Hu,
Phys. Rev. E {\bf 72}, 016211 (2005).

\bibitem{Lind2004}
P.~G. Lind, J.~A.~C. Gallas, and H.~J. Herrmann, Phys. Rev. E {\bf 70}, 056207 (2004).

\bibitem{Atay:2004}
F.~M. Atay, J.~Jost, and A.~Wende,
Phys. Rev. Lett. {\bf 92}, 144101 (2004).

\bibitem{Belykh:2004}
V.~N. Belykh, I.~V. Belykh, and M.~Hasler,
Physica D {\bf 195}, 159 (2004).

\bibitem{Timme:2004}
M.~Timme, F.~Wolf, and T.~Geisel,
Phys. Rev. Lett. {\bf 92}, 074101 (2004).

\bibitem{Wang}
B.~Wang, H.~Tang, T.~Zhou, and Z.~Xiu,
cond-mat/0512079.

\bibitem{Donetti}
L.~Donetti, P.~I. Hurtado, and M.~A. Mu\~noz,
Lect. Notes. Comput. Sc. {\bf 3993}, 1075 (2006).


\bibitem{barahona2002}
M. Barahona and L.~M. Pecora, Phys. Rev. Lett. {\bf 89}, 054101 (2002).

\bibitem{Nishikawa2003}
T. Nishikawa, A.~E. Motter, Y.-C. Lai, and F.~C. Hoppensteadt, Phys. Rev. Lett. {\bf 91}, 014101 (2003).

\bibitem{McGraw2005}
P.~N. McGraw and M. Menzinger, Phys. Rev. E {\bf 72}, 015101(R) (2005).

\bibitem{Kocarev2005}
L.~Kocarev and P.~Amato, Chaos {\bf 15}, 024101 (2005).

\bibitem{Motter2005b}
A.~E. Motter, C.~S. Zhou, and J. Kurths, Europhys. Lett. {\bf 69}, 334 (2005).

\bibitem{Motter2005c}
A.~E. Motter, C.~S. Zhou, and J. Kurths, Phys. Rev. E {\bf 71}, 016116 (2005). 

\bibitem{Motter2005}
A.~E. Motter, C.~S. Zhou, and J. Kurths, AIP Conference Proceedings {\bf 776}, 201 (2005).


\bibitem{Winfree}
A.~T.~Winfree, \textit{The geometry of biological time} (Springer-Verlag, New York, 2001).

\bibitem{Rodriguez}
E.~Rodriguez, N.~George, J.-P.~Lachaux, J.~Martinerie, B.~Renault, and F.~J.~Varela,
Nature {\bf 397}, 430 (1999).

\bibitem{Stopfer}
M.~Stopfer, S.~Bhagavan, B.~H.~Smith, and G.~Laurent,
Nature {\bf 390}, 70 (1997).

\bibitem{Hwang2005}
D.-U. Hwang, M. Chavez, A. Amann, and S. Boccaletti,
Phys. Rev. Lett. {\bf 94}, 138701 (2005).

\bibitem{Nishikawa:2006}
T.~Nishikawa and A.~E. Motter,
Phys. Rev. E {\bf 73}, 065106 (2006).

\bibitem{fishcer:2005}
E. Fischer and U. Sauer, Nature Genetics {\bf 37}, 636 (2005).

\bibitem{Korniss:2003}
C.~Korniss, M.~A. Novotny, H.~Guclu, Z.~Toroczkai, and P.~A. Rikvold,
Science {\bf 299}, 677 (2003).

\bibitem{pecora1998}
L.~M. Pecora and T.~L. Carroll, Phys. Rev. Lett. {\bf 80}, 2109 (1998).

\bibitem{Lu2006}
W. Lu and T. Chen, Physica D {\bf 213}, 214 (2006).

\bibitem{Lu2004a}
W. Lu and T. Chen, 
Physica D {\bf 198}, 148
(2004).

\bibitem{Lu2004b}
W. Lu and T. Chen,
IEEE Trans. Circuit and Syst. I {\bf 51}, 2491 
(2004).

\bibitem{fink2000}
K.~S. Fink, G.~Johnson, T.~Carroll, D.~Mar, and
L.~Pecora, Phys. Rev. E {\bf 61}, 5080 (2000).

\bibitem{Heagy}
J.~F. Heagy, L.~M. Pecora, and T.~L. Carroll,
Phys. Rev. Lett. {\bf 74}, 4185 (1995).

\bibitem{Wu2005}
C.~W.  Wu, Linear Algebra and Its Applications {\bf 402},  207 (2005).

\bibitem{sb}
S. Boccaletti, private communication.

\bibitem{Braess1968}
D. Braess, Unternehmensforschung {\bf 12}, 258 (1968)

\bibitem{Irvine:1993}
A. D. Irvine, Int. Studies in Philosophy of Science {\bf 7}, 145 (1993).

\bibitem{toroczkai2004}
Z.~Toroczkai and K.~E.~Bassler, Nature {\bf 428}, 716 (2004).

\bibitem{Durand}
M.~Durand, 
cond-mat/0608208.

\bibitem{Yook2005}
S.-H. Yook and H. Meyer-Ortmanns, cond-mat/0507422.

\bibitem{Zheng2000}
Z. Zheng, G. Hu, and B. Hu, Phys. Rev. E {\bf 62}, 7501 (2000).

\bibitem{Jing2005}
J.~Jing and K.~R. Weiss,
Current Biology {\bf 15}, 1712 (2005).

\bibitem{Quiroga2005}
R.~Q. Quiroga, L.~Reddy, G.~Kreiman, C.~Koch, and I.~Fried,
Nature {\bf 435}, 1102 (2005).

\end{thebibliography}
\end{document}